\documentclass[12pt]{article}

\usepackage{setspace}
\usepackage{epsfig,lscape,longtable,rotating,graphicx,pdfpages}
\usepackage[titletoc,toc,title]{appendix}
\usepackage{color,soul} % for highlighting
\usepackage{cite}
\usepackage{amsmath}
\newcommand{\nopageno}{\pagestyle{empty}}

\newcommand{\bay}{\begin{eqnarray}}
\newcommand{\eay}{\end{eqnarray}}

\newpage
\begin{document}
\nopageno
\begin{center}
{\Large {\bf Generating survival times using Cox proportional hazards models with cyclic time-varying covariates, with application to a multiple-dose monoclonal antibody clinical trial}}\\
\bigskip

Short title: Generating survival times with cyclic time-varying covariates\\
\bigskip

{\bf Yunda Huang$^{a,b,\ast}$,  Yuanyuan Zhang$^a$, Zong Zhang$^{c}$, Peter B. Gilbert$^{a,d}$}\\
\end{center}

$^a$Vaccine and Infectious Disease Division, Fred Hutchinson Cancer Research Center,  1100 Fairview Ave. N., Seattle, Washington, USA, 98109;\\
$^b$Department of Global Health, University of Washington, 1510 San Juan Rd., Seattle, Washington, USA, 98195; \\
$^c$Interlake High School, 16245 NE 24th St., Bellevue, Washington, USA, 98008;\\
$^d$Department of Biostatistics, University of Washington, 1705 NE Pacific St., Seattle, Washington, USA, 98195;\\
$^\ast$Corresponding author (yunda@fhcrc.org, Tel: 206-667-5780.)

\newpage

\abstract{
In two harmonized efficacy studies to prevent HIV infection through multiple infusions of the monoclonal antibody VRC01, a key objective is to evaluate whether the serum concentration of VRC01, which changes cyclically over time along with the infusion schedule, is associated with the rate of HIV infection. Simulation studies are needed in the development of such survival models. In this paper, we consider simulating event time data with a continuous time-varying covariate whose values vary with time through multiple drug administration cycles, and whose effect on survival changes differently before and after a threshold within each cycle. The latter accommodates settings with a zero-protection biomarker threshold above which the drug provides a varying level of protection depending on the biomarker level, but below which the drug provides no protection. We propose two simulation approaches: one based on simulating survival data under a single-dose regimen first before data are aggregated over multiple doses, and another based on simulating survival data directly under a multiple-dose regimen. We generate time-to-event data following a Cox proportional hazards model based on inverting the cumulative hazard function and a log link function for relating the hazard function to the covariates.
The method's validity is assessed in two sets of simulation experiments. The results indicate that the proposed procedures perform well in producing data that conform to their cyclic nature and assumptions of the Cox proportional hazards model.}
\\
\textbf{keywords:} correlates of risk analysis; exponential distribution; Monte Carlo methods; survival data simulations; time-dependent covariate; zero-protection threshold.  

\section{Introduction}
Time-to-event outcomes with time-varying covariates are frequently encountered in biomedical studies. In the Phase 2b HIV-1 Antibody Mediated Prevention (AMP) efficacy study to prevent HIV-1 infection through ten 8-weekly intravenous infusions of a monoclonal antibody VRC01\cite{Gilbert_SCID2017}, 
participants' drug concentrations in serum are expected to change continuously and cyclically over time, peaking within hours after each infusion, declining at a faster rate in the first few days followed by a decay at a slower rate, and possibly diminishing to an undetectable level through each infusion cycle \cite{LedgerwoodGraham_CEI2015,MayerSeatonHuang2017,HuangZhang_mAbs2017}. Population pharmacokinetics (popPK) analysis based on non-linear mixed effects models is a commonly used tool to estimate population- and individual-level PK parameters that characterize the drug decay process, as well as to estimate drug concentrations over time overall and for each individual. The primary objective of AMP is to evaluate the prevention efficacy of VRC01 (vs. placebo) at dose levels of 10 mg/Kg and 30 mg/Kg. A key secondary objective is to assess the association of VRC01 serum concentration (or other functional biomarkers) over time with the instantaneous rate of HIV infection in the AMP correlates of risk (CoR) analyses \cite{Gilbert_SCID2017}. 

Simulation studies are often needed in the development of such CoR models with time-to-event outcome and time-varying covariates. An essential starting point is to produce simulated survival times from a known data generating process\cite{Leemis_OR1987,Leemis_SPL1990,Bender_SM2005,Bender_SM2006}. For continuous covariates, previous work has been limited to simulating event times with time-varying covariates that follow a simple linear relationship with time and/or log-transformed time\cite{Austin_SM2012,Austin_SM2013,CrowtherLambert_SM2013}, or covariates that change at integer-valued steps of the time scale\cite{Hendry_SM2014} throughout the entire follow up period. Such data generating processes are only appropriate when individuals are uniformly exposed to risk of acquiring the survival outcome at each unit of time (e.g., oral daily dose of the same drug amount). Therefore, new or extensions of these methods are needed for settings like the AMP study, where continuous covariate values change over time in a cyclic non-monotone form.

Cox proportional hazards (PH) regression models are the most common approach for evaluating the effect of covariates, including time-varying covariates on survival outcomes. The objective of this paper is to develop a method for the generation of survival times that follow a Cox PH model with cyclic time-varying covariates. We consider a continuous time-varying covariate whose value varies with time periodically through cycles of multiple drug administrations. In addition, within each cycle, the covariate's effect on survival may differ before and after a threshold value is reached. For example, under a zero-protection threshold model, when above a certain threshold value, the covariate's effect on the survival outcome follows a certain function. However, when below the threshold value, the covariate has no effect on survival. We generate time-to-event data following a Cox PH model based on inverting the cumulative hazard function and a log link function for relating the hazard function to the time-varying and time-invariant covariates. We consider closed-form derivations for simulating three commonly used distributions: Exponential, Weibull and Gompertz, all of which satisfy the PH assumptions. We propose two simulation approaches. The first approach is based on simulating survival data under a single-dose regimen before such data are aggregated over multiple-dose intervals; the second approach is based on simulating survival data directly under a multiple-dose regimen. 

The paper is structured as follows. In Section 2, we briefly review previous work on generating survival times via Cox models. We introduce both the single-dose and multiple-dose approaches for simulating survival time with continuous and cyclic time-varying covariates. For the single-dose approach, we provide, under the zero-protection model, details of the closed-form derivations of event times following an Exponential distribution in the main text; derivations for Weibull- and Gompertz-distributed event times are presented in the Appendix. For the multiple-dose approach, we provide details of the derivations assuming a monotonic relationship between the time-varying covariate and the survival outcome within each dosing cycle in the main text; extensions incorporating the zero-protection model are provided in the Appendix. In Section 3, we describe two simulation experiments to assess the developed method with application to the AMP CoR study. Conclusions are provided in Section 4. 

\section{Simulating survival times}
\subsection{Review}
We first briefly describe basic steps of simulating survival times based on Cox PH model as discussed in\cite{Leemis_OR1987,Leemis_SPL1990,Bender_SM2005,Bender_SM2006,Austin_SM2012,Austin_SM2013,CrowtherLambert_SM2013,Hendry_SM2014}. The Cox PH model is given by
\begin{equation}\label{cox}
h(t|x,z(t)) = h_{0}(t)\mbox{exp}(\beta z(t)+\eta ^\prime x),
\end{equation}
where $z(t)$ denotes the time-varying covariate, whose value changes over the duration of the follow-up time, while its effect on the hazard of the outcome stays constant as denoted by the regression coefficient $\beta$; $x$ denotes the time-invariant covariates, and $\eta$ is the vector of regression coefficients associated with the vector of fixed covariates $x$. $h_{0}(t)$ is the baseline hazard function, i.e., the hazard function of the outcome for those subjects with $x=0$ and $z(t)=0$.

As the Cox model is formulated through the hazard function, the simulation of appropriate survival times for this model needs further manipulation based on the relationship between the hazard function and the covariate. A small number of prior studies have developed methods for simulating event time data with time-varying covariates. Leemis et al. \cite{Leemis_SPL1990} briefly described methods based on inverting the cumulative hazard function to generate event times in settings with time-varying covariates, Sylvestre and Abrahamowicz \cite{Sylvestre_SM2008} described a permutational algorithm and a binomial model for simulating event times conditional of time-varying covariates, and Austin \cite{Austin_SM2012,Austin_SM2013} extended the work of Leemis et al.\cite{Leemis_OR1987,Leemis_SPL1990} using the log link function for relating the hazard function to the linear covariates and incorporated both time-invariant and time-varying covariates.

The translation of the regression coefficients from hazard to survival time is relatively easy if the baseline hazard function is constant, i.e. the survival times are exponentially distributed and $h_{0}(t)=\lambda$, $\lambda>0$. The cumulative hazard function of model (\ref{cox}) is given by:
\begin{equation}
H(t|x,z(t))= \int^{t}_{0}{\lambda\mbox{ exp}(\beta z(u)+\eta^\prime x)\,du}.
\end{equation}
Because the survival function of the above model, $S(t|x,z(t))=\mbox{exp}(-H(t|x,z(t)))$ follows the standard uniform distribution U(0,1), both Leemis and Bender et al. \cite{Leemis_OR1987,Leemis_SPL1990, Bender_SM2005} have demonstrated that a survival time, $T$, can be generated by inverting the cumulative hazard function via $T=H^{-1}(-\mbox{log}(u))$, where $u \sim U(0,1)$. 

\subsection{Proposed methods}
For concreteness, we describe our methods in the context of the AMP study. The same data generating process can be generalized to other applicable biomedical settings, where the association between a time-to-event outcome and a cyclic time-varying covariates is of interest. We define event time, $t$, as time (in days) from study enrollment to HIV-1 infection, and the hazard of HIV-1 infection is modeled as a function of time-varying drug concentration over time according to Equation (1) in a Cox model. Suppose a maximal number of $m$ infusions are planned for each individual. Let $I_1$, $I_2$, ..., $I_{m-1}$ indicate the $m-1$ infusion interval lengths between the $m$ infusions, and $I_m$ indicate the interval between the last infusion and the end of the study. Note that $m$ takes values between 1 and 10 in AMP, and $m$ could differ between individuals due to missed infusions under imperfect infusion adherence. Under a zero-protection threshold model, the time-varying covariate within a given infusion interval $I_k$, $k=1, \ldots, m$, is defined as follows:
\begin{equation}\label{z(t)}
z(t) = 
\begin{cases}
t & \text{if } t\leq t_s,\\
t_s & \textrm{otherwise,}
\end{cases}
\end{equation}
where $t_s$ indicates the time (since infusion) when drug concentration reaches a zero-protection threshold $s$. In other words, for $t \leq t_s$, we consider $z(t)$ as a proxy of drug concentrations at time $t$ because drug concentrations are expected to change with time in a monotonic relationship, and consequently, as shown in Equation (\ref{cox}) the instantaneous hazard $h(t)$ changes at a rate of exp$(\beta)$ per-day change in $t$. For $ t> t_s$, $z_{t}$ remains constant so that $h(t)$ does not continue to change until the next infusion takes place. In many biomedical settings, $t_s$ is considered as the time-point when a therapeutic or protection threshold is achieved. In addition, time-invariant covariates, $x$, could be a vector of individual-level random-effects PK parameters used to describe the inter-individual variability of the PK processes based on non-linear mixed effects modeling of the time-concentration data in a study cohort\cite{HuangZhang_mAbs2017}. An example of $x$ is the estimated individual-level clearance rate of VRC01. We extend the work of Austin \cite{Austin_SM2012} and consider both time-invariant covariates $x$ and a continuous time-varying covariate $z(t)$, whose values change over time in a cyclic form and whose effects on survival change in a piece-wise manner within each cycle. 

Besides the PH assumption, both the single-dose and multiple-dose approaches described below rely on the `cycle-invariant' assumption that the effect of the cyclic time-varying covariate on survival does not change between cycles. This assumption is reasonable in the context of AMP for two reasons. First, pharmacologically steady state is expected to be reached after 5--6 half-lives of a drug. This implies that the time-concentration curve of VRC01 fluctuates in the same pattern over subsequent dosing intervals after the second dose at 8 weeks, because the half life of VRC01 is approximately two weeks. Second, VRC01 exhibits a linear or dose-independent PK feature in that the PK parameters used to describe its time-concentration curve do not change when different doses or multiple doses of VRC01 are given\cite{HuangZhang_mAbs2017}. 

\subsubsection{Single-dose approach}
The single-dose approach considers simulating survival data over one dose interval as a first step before such data are aggregated over multiple dose intervals. Instead of having the same continuous relationship with $t$ throughout the entire follow up time as described in \cite{Austin_SM2013}, $z(t)$ in our case changes at $t_s$ within each drug administration cycle, as shown in Equation (\ref{z(t)}). This feature guards against the possibility of the hazard in the treatment group becoming greater than that in the control group when $t$ gets too large due to a missed infusion. In addition, we assume that the instantaneous hazard at $t_s$ as $h(t=t_s|x,z(t))=\lambda\mbox{ exp}(\beta t_s+\eta^\prime x) = \lambda_p$, or equivalently, $\lambda\mbox{ exp}(\eta^\prime x) = \lambda_{p}\mbox{ exp}(-\beta t_s)$ , where $\lambda_p$ indicates the hazard rate in the control group where no effect of the drug on survival is expected to be observed. 

In reality, $t_s$ could differ across individuals who receive different weight-based dose amounts due to different body weights, or who have different values of other covariates that may influence the inter-individual variability of various PK parameters for VRC01. For simplicity and faster computation, an average $t_s$ can be used in the actual simulation of survival times. For example, based on the popPK model of VRC01 described in \cite{HuangZhang_mAbs2017}, we estimate that it takes an average of 57 or 81 days, respectively, for the drug concentration of a potential AMP participant with body weight of 74.5 Kg receiving the 10mg/Kg or 30 mg/Kg dose VRC01 to decline to $s=5.0$ mcg/mL, a VRC01 concentration that is hypothesized to confer protection against HIV infection \cite{PeguNabelSCM2014,RudicellJV2014,KoNature2014,SaundersJV2015}. This implies that $t_s=57$ days for the 10 mg/Kg dose group, and $t_s=81$ days for the 30 mg/Kg dose group. The instantaneous hazard remains constant after 57 and 81 days, respectively, in the low and high dose groups. 
%we estimate that it takes an average of 90 or 114 days, respectively, for the drug concentration of a potential AMP participant with body weight of 74.5 Kg receiving the 10mg/Kg or 30 mg/Kg dose VRC01 to decline to $s=1.0$ mcg/mL, a lower limit of hypothetical VRC01 concentration levels that may confer protection against HIV infection\cite{PeguNabelSCM2014,RudicellJV2014,KoNature2014,SaundersJV2015}. This implies that $t_s=90$ days for the 10 mg/Kg dose group, and $t_s=114$ days for the 10 mg/Kg dose group. The instantaneous hazard remains constant after 90 and 120 days, respectively, in the low and high dose groups. 
This ensures meaningful simulated survival time to account for the wide infusion visit window in AMP (-1 week to +7 weeks around the target 8-weekly infusion visits) and for individuals whose infusion intervals are great than 8 weeks due to missed infusions.

Now, we describe first the steps to simulate survival times after a single dose, by inverting the cumulative hazard function. In the following, we show derivations in details for Exponential distribution of survival times; details for the Weibull and Gompertz distributions are reported in Appendix A1 and A2, respectively. 

For exponentially-distributed survival times, $h_{0}(t)=\lambda$.  
If $t \le t_s$, the cumulative hazard function is equal to
\begin{eqnarray}\nonumber
H(t,x,z(t))&=& \int^{t}_{0}{\lambda\mbox{ exp}(\beta z(u)+\eta^\prime x)\,du} \nonumber\\
 &=& \int^{t}_{0}{\lambda\mbox{ exp}(\beta u+\eta^\prime x)\,du} \nonumber\\
 &=& \lambda\mbox{ exp}(\eta^\prime x)\int^{t}_{0}{\mbox{ exp}(\beta u)\,du} \nonumber\\
 &=& \lambda\mbox{ exp}(\eta^\prime x)\left[\frac{1}{\beta}\mbox{ exp}(\beta u)\right]_{0}^{t} \nonumber\\
 &=& \frac{\lambda\mbox{ exp}(\eta^\prime x)}{\beta}\left[\mbox{ exp}(\beta t)-1\right].\nonumber
\end{eqnarray}

Consequently, the inverse cumulative hazard function is
\begin{equation*}
H^{-1}(v) = \frac{1}{\beta}\mbox{ log}\left(1+\frac{\beta v}{\lambda\mbox{ exp}(\eta^\prime x)} \right).
\end{equation*}
Therefore, an event time can be generated as
\begin{equation}
T = \frac{1}{\beta}\mbox{ log}\left(1+\frac{\beta (-\mbox{ log}(u))}{\lambda\mbox{ exp}(\eta^\prime x)} \right)
\mbox{, if} -\mbox{log}(u) < \frac{\lambda\mbox{ exp}(\eta^\prime x)}{\beta}\left[\mbox{ exp}(\beta t_s)-1\right],
\end{equation}
where $u\sim U(0,1)$.

If $t > t_s$, the cumulative hazard function is equal to
\begin{eqnarray}
H(t,x,z(t)) &=& \int^{t_s}_{0}{\lambda \mbox{ }(\beta u + \eta^\prime x)\,du}+ \int^{t}_{t_s}{
   \lambda \mbox{ exp}(\beta t_s + \eta^\prime x),du} \nonumber\\
 &=& \lambda\mbox{ exp}(\eta^\prime x)\left(\frac{1}{\beta}(\mbox{exp}(\beta t_s)-1)
  + (t-t_s)\mbox{ exp}(\beta t_s)\right).\nonumber
\end{eqnarray}

Consequently, the inverse cumulative hazard function is
\begin{equation*}
H^{-1}(v) = \frac{v}{\lambda\mbox{ exp}(\beta t_s +\eta^\prime x)} + \frac{1-\mbox{ exp}(\beta t_s)}{\beta \mbox{ exp}(\beta t_s)} + t_s.
\end{equation*}
Therefore, an event time can be generated as
\begin{multline}
T = \frac{-\mbox{log}(u)}{\lambda\mbox{ exp}(\beta t_s +\eta^\prime x)} + \frac{1-\mbox{ exp}(\beta t_s)}{\beta \mbox{ exp}(\beta t_s)} + t_s,
\mbox{ if } -\mbox{log}(u) \ge \frac{\lambda\mbox{ exp}(\eta^\prime x)}{\beta}\left[\mbox{ exp}(\beta t_s)-1\right],
\end{multline}
%\begin{equation*}
%-\mbox{log}(u) \ge \frac{\lambda\mbox{ exp}(\beta^\prime x)}{\beta_{t}}\left[\mbox{ exp}(\beta_{t}t_s)-1\right]
%\end{equation*}
where $u\sim U(0,1)$.

In summary, in order to simulate survival times under a zero-protection threshold model after a single dose is given, a random uniform sample, $u$ is first simulated and the survival time takes the form in Equation (4) if $-\mbox{log}(u) < \frac{\lambda\mbox{ exp}(\eta^\prime x)}{\beta}\left[\mbox{ exp}(\beta t_s)-1\right]$, or the form in Equation (5), otherwise.

After the single-dose survival time according to the Exponential, Weibull, or Gompertz distribution is simulated as described above or in the Appendix, the survival time after multiple doses can be simulated as follows:
\begin{itemize}
\item[1.] Simulate the infusion times for each individual's $m$ infusions. Infusion visit windows and possible missed infusions and/or permanent infusion discontinuations could be considered here; 
\item[2.] For each individual, independently simulate the single-dose survival time $T_1$, $T_2$, ..., $T_m$ for each of the $m$ infusion intervals according to equations [4] and [5]; 
\item[3.] If all $T_k> I_k$, $k=1$, 2, \ldots, $m$, then the final multiple-dose survival time of this uninfected individual is censored at $S=\sum\limits_{i=1}^{m}I_i$. Otherwise, pick the first $k$ that satisfies $T_k < I_k$, and the final multiple-dose survival time for this infected individual is $ S = \sum\limits_{i=1}^{k-1}I_i + T_k$.
\end{itemize}

This approach guarantees that, as desired, the event time follows the same survival function within each infusion interval, and the probability of infection during a given interval is not affected by the probability of the same individual not being infected in the prior infusion interval because P(infected in $t_2$) = P(infected in $t_2$ $|$ not infected in $t_1$) = P ($T_2< t_2$ $|$ $T_1> t_1$) = P ($T_2 < t_2$) (given the all $T_k$'s are i.i.d).

\subsubsection{Multiple-dose approach}
The multiple-dose approach considers simulating survival data over multiple dose intervals directly. In the multiple-dose setting, let $(t_1,...,t_m)$ denote the actual infusion time (since enrollment) for the first to last $m^{th}$ infusion one receives, where $m \le 10$ according to the AMP protocol. The first infusion coincides with enrollment and hence $t_1=0$. Using the notations from the single-dose approach, $t_{i+1}=t_{i} + I_{i}$, for $i=1, 2, ..., m\textrm{-1}$.

If $t_s$ is always greater than all dosing intervals, e.g., under perfect adherence to the 8-weekly infusion schedule, the following steps can be used to generate survival times for participants receiving up to $m$ doses. If $t_s$ may be smaller than a dosing interval, then similar strategies as illustrated in Section 2.1.1 by combining the cumulative hazards before $t_s$ and after $t_s$ can be employed for simulating survival times via the multiple-dose approach (Appendix: A3). Similarly, survival times can be simulated by inverting the cumulative hazard function. In the following derivations, survival times are assumed to be exponentially-distributed.

If $t_1 \le t<t_2$, the cumulative hazard function is equal to
\begin{eqnarray}
H(t,x,z(t))&=& \int^{t}_{0}{\lambda\mbox{ exp}(\beta z(u)+\eta^\prime x)\,du} \nonumber\\
 &=& \int^{t}_{0}{\lambda\mbox{ exp}(\beta u+\eta^\prime x)\,du} \nonumber\\
 &=& \lambda\mbox{ exp}(\eta^\prime x)\int^{t}_{0}{\mbox{ exp}(\beta u)\,du} \nonumber\\
 &=& \lambda\mbox{ exp}(\eta^\prime x)\left[\frac{1}{\beta}\mbox{ exp}(\beta u)\right]_{0}^{t} \nonumber\\
 &=& \frac{\lambda}{\beta}\mbox{ exp}(\eta^\prime x)\left[\mbox{ exp}(\beta t)-1\right]. \nonumber
\end{eqnarray}

Consequently, the inverse cumulative hazard function is
\begin{equation} \nonumber
H^{-1}(u) = \frac{1}{\beta}\mbox{ log}\left(1+\frac{\beta u}{\lambda\mbox{ exp}(\eta^\prime x)} \right).
\end{equation}
Therefore, an event time can be generated as
\begin{equation}
T = \frac{1}{\beta}\mbox{ log}\left(1+\frac{\beta (-\mbox{ log}(u))}{\lambda\mbox{ exp}(\eta^\prime x)} \right)
\mbox{, if} -\mbox{log}(u) < b_1 
\end{equation}
where $b_1 = \frac{\lambda}{\beta}\mbox{ exp}(\eta^\prime x)\left[\mbox{ exp}(\beta t_2)-1\right],$ and $u\sim U(0,1)$.\\

If $t_2\le t < t_3$, the cumulative hazard function is equal to
\begin{eqnarray}
H(t,x,z(t))&=& \int^{t}_{0}{\lambda\mbox{ exp}(\beta z(u)+\eta^\prime x)\,du} \nonumber\\
 &=& \lambda\mbox{ exp}(\eta^\prime x)\int^{t}_{0}{\mbox{exp}(\beta u)\,du} \nonumber\\
 &=& \lambda\mbox{ exp}(\eta^\prime x)\left(\int^{t_2}_{0}{\mbox{exp}(\beta u)\,du}+ \int^{t}_{t_2}
   {\mbox{ exp}(\beta (u-t_2))\,du}\right) \nonumber\\
 &=& \lambda\mbox{ exp}(\eta^\prime x)\left(\frac{1}{\beta}(\mbox{exp}(\beta t_2)-1)
   + \frac{1}{\beta}(\mbox{exp}(\beta t-\beta t_2)-1)\right)\nonumber\\
 &=& \frac{\lambda}{\beta}\mbox{exp}(\eta^\prime x)\left(\mbox{exp}(\beta t_2)
   + \mbox{exp}(\beta t-\beta t_2)-2\right).\nonumber
\end{eqnarray}

Consequently, the inverse cumulative hazard function is
\begin{equation} \nonumber
H^{-1}(u) = \frac{1}{\beta}\mbox{ log}\left(\mbox{exp}(\beta t_2)\left( \frac{\beta u}{\lambda \mbox{
exp}(\eta^\prime x)}-\mbox{exp}(\beta t_2)+2\right) \right).
\end{equation}
Therefore, an event time can be generated as
\begin{equation}\nonumber
T = \frac{1}{\beta}\mbox{ log}\left(\mbox{exp}(\beta t_2)\left( \frac{\beta (-\mbox{log}(u))}{\lambda \mbox{
exp}(\eta^\prime x)}-\mbox{exp}(\beta t_2)+2\right) \right)\mbox{, if } a_2 \le -\mbox{log}(u) < b_2,
\end{equation}
where
\begin{eqnarray}
a_2 &=& \frac{\lambda}{\beta}\mbox{ exp}(\eta^\prime x)\left(\mbox{ exp}(\beta t_2)-1\right),\nonumber\\
b_2 &=&  \frac{\lambda}{\beta}\mbox{exp}(\eta^\prime x)\left(\mbox{exp}(\beta t_2)+ \mbox{exp}(\beta t_3-\beta t_2)-2\right),\text{ and}\nonumber\\
u &\sim& U(0,1).\nonumber
\end{eqnarray}

Similarly, for $t_{k}\le t < t_{k+1}$, $k=2, \ldots, m-1$, the cumulative hazard function is equal to
\begin{equation}\label{cumhaz}
H(t,x,z(t)) = \frac{\lambda}{\beta}\mbox{exp}(\eta^\prime x)\left[\sum\limits_{i=2}^{k}\mbox{exp}(\beta (t_{i}-t_{i-1}))
   + \mbox{exp}(\beta t-\beta t_{k})-k\right].\nonumber
\end{equation}
And the inverse cumulative hazard function is
\begin{equation*}
H^{-1}(u) = \frac{1}{\beta}\mbox{ log}\left(\mbox{exp}(\beta t_{k})\left( \frac{\beta u}{\lambda \mbox{exp}(\eta^\prime x)}-\sum\limits_{i=2}^{k}\mbox{exp}(\beta (t_{i}-t_{i-1}))+k\right)\right).
\end{equation*}

Therefore, an event time can be generated as
\begin{equation}
T=\frac{1}{\beta}\mbox{ log}\left(\mbox{exp}(\beta t_{k})\left(\frac{\beta (-\mbox{log}(u))}{\lambda \mbox{exp}(\eta^\prime x)}-\sum\limits_{i=2}^{k}\mbox{exp}(\beta (t_{i}-t_{i-1}))+k\right)\right)\mbox{, if } a_k \le -\mbox{log}(u) < b_k,
\end{equation}
where,
\begin{eqnarray} \nonumber
a_k &=& \frac{\lambda}{\beta}\mbox{ exp}(\eta^\prime x)\left(\sum\limits_{i=2}^{k}\mbox{ exp}(\beta (t_{i}-t_{i-1}))-(k-1)\right),\\
b_k &=& \frac{\lambda}{\beta}\mbox{ exp}(\eta^\prime x)\left(\sum\limits_{i=2}^{k}\mbox{ exp}(\beta (t_{i}-t_{i-1})+ \mbox{exp}(\beta t_{k+1}-\beta t_{k})-k\right), \text{ and}\nonumber\\
u &\sim& U(0,1).\nonumber
\end{eqnarray}

Lastly, if $t\ge t_m$, an event time can be generated as
\begin{equation}
T=\frac{1}{\beta}\mbox{ log}\left(\mbox{exp}(\beta t_{m})\left(\frac{\beta (-\mbox{log}(u))}{\lambda \mbox{
exp}(\eta^\prime x)}-\sum\limits_{i=2}^{m}\mbox{exp}(\beta (t_{i}-t_{i-1})+m\right)\right)\mbox{, if }-\mbox{log}(u) \ge b_m,
\end{equation}
where $b_m = \frac{\lambda}{\beta}\mbox{exp}(\eta^\prime x)\left(\sum\limits_{i=2}^{m}\mbox{ exp}(\beta (t_{i}-t_{i-1})-(m-1))\right)$ and $u\sim U(0,1)$.\\

In summary, in order to simulate survival times in a multiple-dose setting, the infusion times $(t_1,...,t_m)$ according to the study set-up and a random uniform sample, $u$ are first simulated. Then, for each $k=1, \ldots, m-1$, $a_k$ and $b_k$ are calculated where $a_1$=0 and $a_{k+1}=b_k$. The survival time takes the form in Equation (6) if $-\mbox{log}(u) < b_1$, or the form in Equation (7) if $a_k \le -\mbox{log}(u) < b_k$, or the form in Equation (8) if $-\mbox{log}(u) \ge b_9$.

\section{Examples}
We next illustrate the validity of the described survival data generating processes in two simulation experiments. In the first experiment, the single-dose approach is used to simulate survival data for 1000 AMP-like trials, each with $n=1500$ participants in each of the 10 mg/Kg VRC01, 30 mg/Kg VRC01 and placebo groups. Within each trial, the time-varying covariate (i.e., drug concentration over time) is associated with the survival outcome (i.e., time to HIV infection in days) according to Equation ~\eqref{cox} with $\beta=0.03$ and $\eta=0$ for both dose groups, and $h_{0}(t)=0.04/365/\textrm{exp}(\beta t_s)$, where $t_s=57$ and $t_s=81$ for the low and high dose groups, respectively to ensure the same baseline HIV infection rate beyond $t_s$ in the two dose groups. In addition, $z(t)$ takes the piece-wise form as described in Equation (\ref{z(t)}) with a zero-protection threshold $s=5$ mcg/mL. These parameter values indicate that, before an individual's drug concentration reaches 5 mcg/mL, the hazard ratio over a 28-day period is exp$(28*0.03)=2.31$, but the rate of infection remains constant ($=0.04$/year) once the individual's drug concentration falls below 5 mcg/mL. We consider two study adherence levels: the high and medium adherence scenarios assume 2\% and 10\% of infusion visits missed, respectively. Consequently, we expect three patterns in the simulated data. First, the low dose group should have higher risk of infection than the high dose group. This is because drug concentrations in the former group on average are expected to reach the zero-protection threshold, 5 mcg/mL in a shorter time or, in another word, the lower dose group is expected to have a smaller $t_{s=5 mcg/mL}$ than the higher dose group, although the two dose groups do have the same risk (due to having the same $\beta=0.03$) until their respective $t_{s=5 mcg/mL}$ time-points within each dosing cycle. Second, a lower risk of infection should be associated with a better study adherence due to less missed infusions and less follow up time with concentration below the zero-protection threshold $s=5$ mcg/mL. Third, a shorter duration between time of infection and prior infusion should occur with better study adherence due to shorter average infusion intervals when there are less missed infusions, although a smaller number of infections do occur with a better study adherence. As shown in Figure 1, all these patterns are confirmed. In addition, the mean (standard deviation) values of the estimated $\beta$ from fitting the simulated data in a standard Cox model with $z(t)=$ time since prior infusion as the time-varying covariate and $t_s \ge$ study duration are 0.021(0.007) and 0.025 (0.01), respectively, for the low and high dose groups under high adherence. Under medium adherence, these values are 0.013 (0.004) and 0.018 (0.006), respectively, for the low and high dose groups. We note that under each adherence scenario, both $\hat{\beta}$ values are smaller than the true value of $\beta=0.03$ because by setting $t_s$ greater, the unit-effect of $z(t)$ is expected to be smaller to achieve the same cumulative effect of $z(t)$ on $h(t)$. We also note that under both adherence scenarios, $\hat{\beta}$ is greater in the high dose group than in the low dose group because in the simulated datasets the high dose group subjects are more protected due to a longer time for their concentration to drop to $s$ and hence a larger unit-effect of $z(t)$ is needed. Lastly, as expected, $\hat{\beta}$ values get closer to 0.03 as adherence improves.

In the second experiment, the multiple-dose approach is used to simulate AMP-like trials under perfect study adherence scenarios with $\eta=0$ and $h_{0}(t)=0.04/365/\textrm{exp}(\beta*56)$. Each trial includes $n=1500$ VRC01 recipients in each of the 10 mg/Kg and 30 mg/Kg dose groups. The same $\beta$ value is used for both dose groups, but two different $\beta$ values: 0.01 and 0.03 are considered in order to verify how risk of infection varies by $\beta$. Figure 2 shows that the probability of HIV infection within each 8-weekly infusion cycle is smaller as $\beta$ gets larger. This pattern is also expected because a higher $\beta$ indicates a larger effect of the biomarker in reducing the risk of infection. In addition, as desired, the rate of HIV infection increases over time (as concentration gets lower) within each infusion cycle, and the pattern remains the same over all cycles under the `cycle-invariant' assumption described in Section 2.2.

\section{Conclusions}
In this paper, we considered simulating event time data with a continuous time-varying covariate whose values vary with time through multiple repetitive cycles, and whose effect on survival changes differently before and after a threshold within each cycle. The latter particularly accommodates settings with a zero-protection biomarker threshold above which the drug provides a varying level of protection depending on the biomarker level, but below which the drug provides no protection. We proposed two simulation approaches: one based on simulating survival data under a single-dose regimen first before data are aggregated over multiple doses, and another based on simulating survival data directly under a multiple-dose regimen. The derivations of the former are more straightforward for handling different event time distributions and can be more easily extended to data models with multiple protection threshold values within a cycle. The derivations of the latter are more compact and simulations based on the latter approach are generally faster than those based on the former approach. The latter approach is also more flexible to be extended to data model where different $z(t)$ functions may be needed for different drug administration cycles. 

The validity of our proposed methods were assessed in two sets of simulation experiments. The results indicate that the proposed procedures perform well in producing data that conform to their cyclic nature and the assumptions of the Cox PH model. Extension can be considered to add the number of doses as another time-dependent covariate. Consequently, the `cycle-invariant' assumption about the effect of the time-varying covariates not changing between cycles can hence be relaxed. Lastly, for drugs that do not satisfy the `cycle-invariant' assumption, different $\beta$ coefficients can be assumed for each cycle and derivations of the simulation procedure based on the multiple-dose approach can be similarly extended for such data models. 

\section*{Disclosure statement}
No potential conflicts of interest were disclosed. 

\section*{Funding}
This work was supported by the National Institute of Allergy and Infectious Diseases (NIAID) US. Public Health Service Grant UM1 AI068635 [HVTN SDMC FHCRC]. The content of this manuscript is solely the responsibility of the authors and does
not necessarily represent the official views of the National Institutes of Health. The funders had no role in study design, data collection and analysis, decision to publish, or preparation of the manuscript.

%\bibliography{PKsimulation}{}
%\bibliographystyle{plain}

\newpage

\textbf {Figure 1: Distributions of simulated event times since prior infusion (Panel A) and since the first infusion (Panel B) under imperfect study adherences.} In these simulations, an annual HIV incidence rate of 4\% is assumed for the placebo group and $\beta=0.03$ or HR$=2.32$ per-28 days for both dose groups with the zero-protection concentration threshold $s=$ 5 mcg/mL in simulated trials of 4500 participants with a 1:1:1 ratio for the three treatment groups. 
\begin{center}
\includegraphics[width=0.8\textwidth]{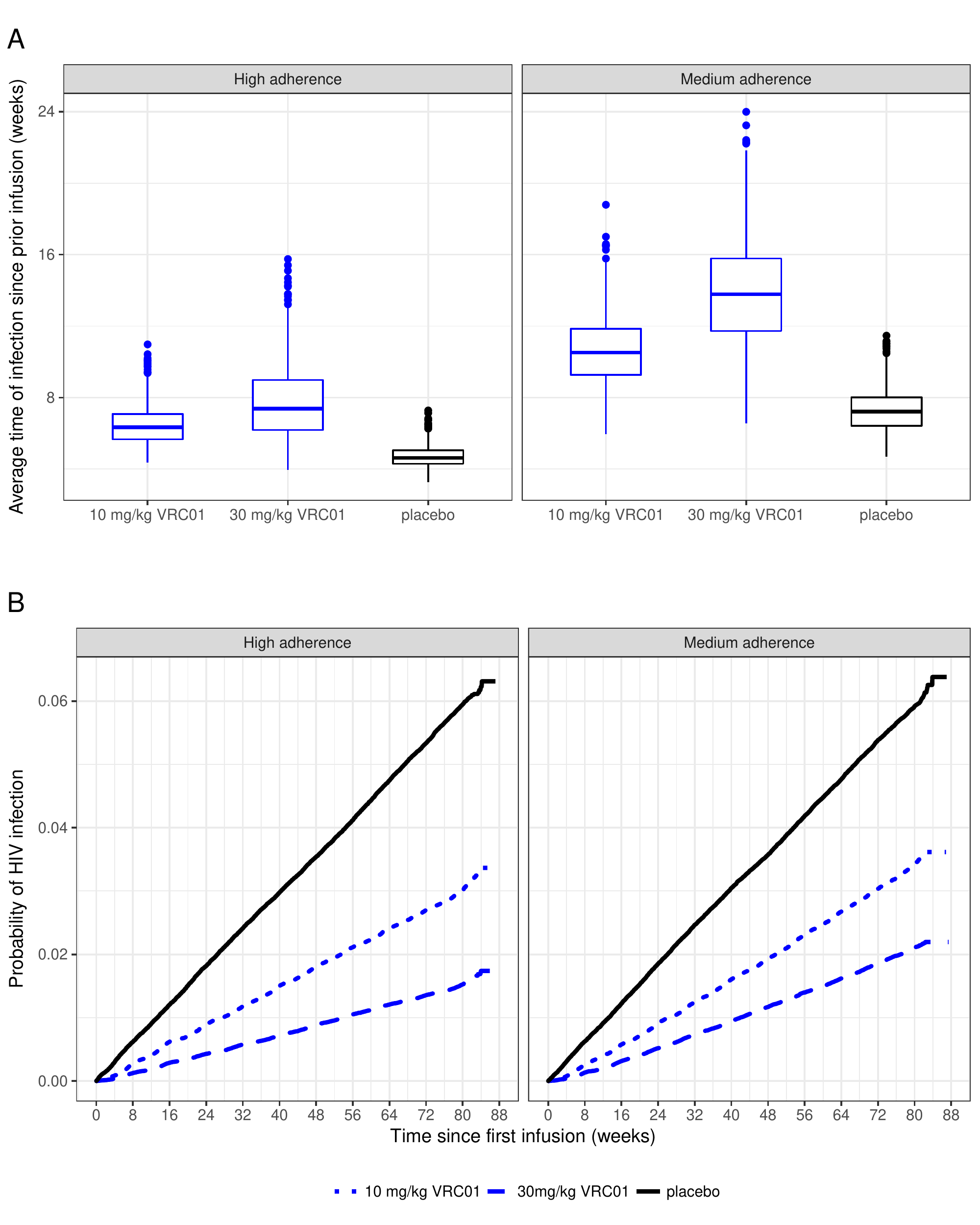}
\end{center}

\newpage
\textbf {Figure 2: Probability of HIV infection within each infusion interval following ten 8-weekly IV infusions of VRC01 under perfect study adherence in a simulated trial of 3000 VRC01 recipients.} Red lines are for $\beta=0.01$ or $HR=1.32$ per-28 days; blue lines are for $\beta=0.03$ or HR$=2.32$ per-28 days. 
\begin{center}
\includegraphics[width=1.0\textwidth]{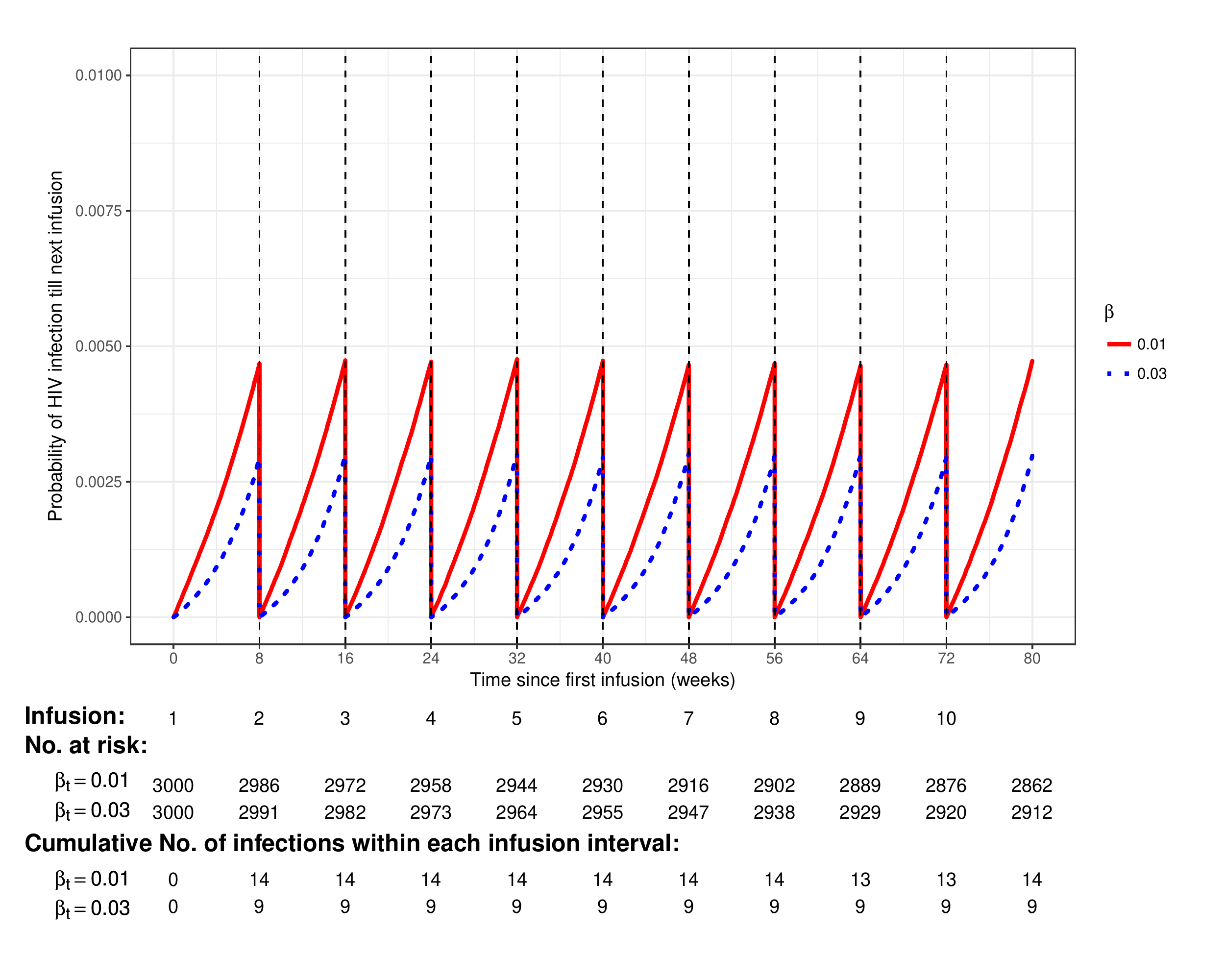}
\end{center}

\newpage
%\includepdf[pages={1-}, landscape=true]{cuminc.pdf}
\begin{center}
{\bf Appendix}
\end{center}

\noindent{\bf A1: single-dose approach assuming Weibull distribution of survival times}\\
The log of the Weibull hazard function is linear in log(t) and the hazard function can be written as 
$h_{0}(t) = \lambda\gamma t^{\gamma -1}$. Hence, the cumulative hazard function becomes
$$H(t|x,z(t))= \int^{t}_{0}{\lambda\gamma u^{\gamma -1}\mbox{ exp}(\beta z(u)+\eta^\prime x)\,du}
.$$
If $t \le t_s$, the cumulative hazard function is equal to
\begin{eqnarray}
H(t,x,z(t))&=& \int^{t}_{0}{\lambda\gamma u^{\gamma -1}\mbox{ exp}(\beta z(u)+\eta^\prime x)\,du} \nonumber\\
 &=& \int^{t}_{0}{\lambda\gamma u^{\gamma -1}\mbox{ exp}(\beta u+\eta^\prime x)\,du} \nonumber\\
 &=& \lambda\gamma\mbox{ exp}(\eta^\prime x)\int^{t}_{0}{u^{\gamma -1}\mbox{ exp}(\beta u)\,du} \nonumber\\
 &=& \lambda\gamma\mbox{exp}(\eta^\prime x)\left[(-\beta)^{-\gamma} \Gamma \left( \gamma, -\beta u \right)\right]_{0}^{t} \nonumber\\
 &=& \frac{\lambda\gamma\mbox{ exp}(\eta^\prime x)}{(-\beta)^{\gamma}}\left[ \Gamma \left( \gamma, -\beta t \right) - \Gamma \left( \gamma, 0\right)\right]\nonumber
\end{eqnarray}

Consequently, the inverse cumulative hazard function is
\begin{equation*}
H^{-1}(v) = -\frac{1}{\beta}\Gamma^{-1}\Bigg[\gamma, \frac{(-\beta)^{\gamma} v}{\lambda\mbox{ exp}(\eta^\prime x)} + \Gamma(\gamma, 0)\Bigg]
\end{equation*}
where $\Gamma^{-1}(\gamma, f(x))$ represents the inverse upper incomplete gamma function. Therefore, an event time can be generated as
\begin{multline}
T = -\frac{1}{\beta}\Gamma^{-1}\Bigg[\gamma, \frac{(-\beta)^{\gamma} (-\log(u))}{\lambda\mbox{ exp}(\eta^\prime x)} + \Gamma(\gamma, 0)\Bigg]\mbox{, if} \\
-\mbox{log}(u) < \frac{\lambda\gamma\mbox{ exp}(\eta^\prime x)}{(-\beta)^{\gamma}}\left[ \Gamma \left( \gamma, -\beta t_{s} \right) - \Gamma \left( \gamma, 0\right)\right]
\end{multline}
where $u\sim U(0,1)$.\\

If $t > t_s$, the cumulative hazard function is equal to
\begin{eqnarray}
H(t,x,z(t)) &=& \int^{t_s}_{0}{\lambda\gamma u^{\gamma -1}\mbox{ exp}(\beta u+\eta^\prime x)du}+ \int^{t}_{t_s}{\lambda\gamma t_{s}^{\gamma -1}\mbox{ exp}(\beta t_{s}+\eta^\prime x)du} \nonumber\\
 &=& \lambda\gamma\mbox{ exp}(\eta^\prime x)\left(\frac{1}{(-\beta)^{\gamma}}( \Gamma \left( \gamma, -\beta t_{s} \right) - \Gamma \left( \gamma, 0\right))
  + (t-t_s)t_{s}^{\gamma-1}\mbox{ exp}(\beta t_s)\right)\nonumber
\end{eqnarray}

Consequently, the inverse cumulative hazard function is
\begin{equation*}
H^{-1}(v) = \frac{v}{t_{s}^{\gamma-1}\lambda\gamma\mbox{ exp}(\beta t_s +\eta^\prime x)} - \frac{\Gamma \left( \gamma, -\beta t_{s} \right) - \Gamma \left( \gamma, 0\right)}{t_{s}^{\gamma-1}(-\beta)^{\gamma}\mbox{ exp}(\beta t_s)} + t_s.
\end{equation*}
Therefore, an event time can be generated as
\begin{multline}
T = \frac{-\log(u)}{t_{s}^{\gamma-1}\lambda\gamma\mbox{ exp}(\beta t_s +\eta^\prime x)} - \frac{\Gamma \left( \gamma, -\beta t_{s} \right) - \Gamma \left( \gamma, 0\right)}{t_{s}^{\gamma-1}(-\beta)^{\gamma}\mbox{ exp}(\beta t_s)} + t_s \mbox{, if }\\
-\mbox{log}(u) \ge \frac{\lambda\gamma\mbox{ exp}(\eta^\prime x)}{(-\beta)^{\gamma}}\left[ \Gamma \left( \gamma, -\beta t_{s} \right) - \Gamma \left( \gamma, 0\right)\right]
\end{multline}

%\begin{equation*}
%-\mbox{log}(u) \ge \frac{\lambda\mbox{ exp}(\eta^\prime x)}{\beta}\left[\mbox{ exp}(\betat_s)-1\right]
%\end{equation*}
where $u\sim U(0,1)$.\\

\noindent
{\bf A2: single-dose approach assuming Gompertz distribution of survival times}

The log of the Gompertz hazard function is linear in t and the hazard function can be written as $h_{0}(t) = \lambda\mbox{ exp}(\alpha t)$. Hence, the cumulative hazard function becomes
$$H(t|x,z(t))= \int^{t}_{0}{\lambda\mbox{ exp}(\alpha u)\mbox{ exp}(\beta z(u)+\eta^\prime x)\,du}.$$

If $t \le t_s$, the cumulative hazard function is equal to
\begin{eqnarray}
H(t,x,z(t))&=& \int^{t}_{0}{\lambda\mbox{ exp}(\alpha u)\mbox{ exp}(\beta z(u)+\eta^\prime x)\,du} \nonumber\\
 &=& \int^{t}_{0}{\lambda\mbox{ exp}(\alpha u)\mbox{ exp}(\beta u+\eta^\prime x)\,du} \nonumber\\
 &=& \lambda\mbox{ exp}(\eta^\prime x)\int^{t}_{0}{\mbox{ exp}((\beta + \alpha)u)\,du} \nonumber\\
 &=& \lambda\mbox{ exp}(\eta^\prime x)\left[\frac{1}{\beta + \alpha}\mbox{ exp}((\beta + \alpha)u)\right]_{0}^{t} \nonumber\\
 &=& \frac{\lambda\mbox{ exp}(\eta^\prime x)}{\beta + \alpha}\left[\mbox{ exp}((\beta + \alpha)t)-1\right]\nonumber
\end{eqnarray}

Consequently, the inverse cumulative hazard function is
\begin{equation*}
H^{-1}(v) = \frac{1}{\beta + \alpha}\mbox{ log}\left(1+\frac{(\beta + \alpha) v}{\lambda\mbox{ exp}(\eta^\prime x)} \right).
\end{equation*}
Therefore, an event time can be generated as
\begin{multline}
T = \frac{1}{\beta + \alpha}\mbox{ log}\left(1+\frac{(\beta + \alpha) (-\mbox{ log}(u))}{\lambda\mbox{ exp}(\eta^\prime x)} \right)\mbox{, if} \\
-\mbox{log}(u) < \frac{\lambda\mbox{ exp}(\eta^\prime x)}{\beta + \alpha}\left[\mbox{ exp}((\beta + \alpha)t_s)-1\right]
\end{multline}
where $u\sim U(0,1)$.\\

If $t > t_s$, the cumulative hazard function is equal to
\begin{eqnarray}
H(t,x,z(t)) &=& \int^{t_s}_{0}{\lambda \mbox{ }((\beta + \alpha)u + \eta^\prime x)\,du}+ \int^{t}_{t_s}{
   \lambda \mbox{ exp}((\beta + \alpha)t_s + \eta^\prime x),du} \nonumber\\
 &=& \lambda\mbox{ exp}(\eta^\prime x)\left(\frac{1}{\beta + \alpha}(\mbox{exp}((\beta + \alpha)t_s)-1)
  + (t-t_s)\mbox{ exp}((\beta + \alpha)t_s)\right)\nonumber
\end{eqnarray}

Consequently, the inverse cumulative hazard function is
\begin{equation*}
H^{-1}(v) = \frac{v}{\lambda\mbox{ exp}((\beta + \alpha)t_s +\eta^\prime x)} + \frac{1-\mbox{ exp}((\beta + \alpha)t_s)}{(\beta + \alpha)\mbox{ exp}((\beta + \alpha)t_s)} + t_s.
\end{equation*}
Therefore, an event time can be generated as
\begin{multline}
T = \frac{-\mbox{log}(u)}{\lambda\mbox{ exp}((\beta + \alpha)t_s +\eta^\prime x)} + \frac{1-\mbox{ exp}((\beta + \alpha)t_s)}{(\beta + \alpha)\mbox{ exp}((\beta + \alpha)t_s)} + t_s \mbox{, if } \\
-\mbox{log}(u) \ge \frac{\lambda\mbox{ exp}(\eta^\prime x)}{\beta + \alpha}\left[\mbox{ exp}((\beta + \alpha)t_s)-1\right]
\end{multline}
%\begin{equation*}
%-\mbox{log}(u) \ge \frac{\lambda\mbox{ exp}(\eta^\prime x)}{\beta}\left[\mbox{ exp}(\betat_s)-1\right]
%\end{equation*}
where $u\sim U(0,1)$.\\

\noindent
{\bf A3: multiple-dose approach assuming imperfect infusion adherence}

In a multiple-dose setting, perfect adherence to the 8-weekly infusion schedule is not always assured. If the ``zero-protection'' threshold $ t_{s} $ is smaller than a dosing interval, then modifications of the derivations covered in Section 2.2.2 are needed when the next infusion occurs after $t_{s}$ has passed.

As stated in Section 2.2.1, in a single-dose setting, for $t>t_{s}$, the cumulative hazard function is equal to
$$H(t, x, z(t)) = \frac{\lambda\exp(\eta'x)}{\beta}[\exp(\beta t) - 1] + \lambda\exp(\eta'x)(t - t_{s})\exp(\beta t_{s}).$$ And, an event time can be generated as 
\begin{multline}
T = \frac{-\log(u)}{\lambda\exp(\beta t_{s} + \eta'x)} + \frac{1-\exp(\beta t_{s})}{\beta\exp(\beta t_{s})} + t_{s}\mbox{, if}\\
-\log(u) \geq \frac{\lambda\exp(\eta'x)}{\beta}[\exp(\beta t_{s}) -1],
\end{multline}
where $u \sim U(0, 1)$.

As stated in Section 2.2.2, in a multiple-dose setting, for $t_{k} \leq t < t_{k+1}, k = 1...., m-1,$ the cumulative hazard function is equal to 
$$H(t, x, z(t)) = \frac{\lambda}{\beta}\exp(\eta'x)\Bigg[\sum_{i=2}^{k} \exp(\beta(t_{i}-t_{i-1})) + \exp(\beta t - \beta t_{k}) - k\Bigg].$$ And, the event time can be generated as 
\begin{multline}
T = \frac{1}{\beta}\log\Bigg(\exp(\beta t_{k})\Bigg(\frac{\beta(-\log(u))}{\lambda\exp(\eta'x)} - \sum_{i=2}^{k} \exp(\beta(t_{i} - t_{i-1})) + k \Bigg)\Bigg)\mbox{, if}\\ a \leq -\log(u) < b,
\end{multline}
where,
\begin{equation*}
a = \frac{\lambda}{\beta}\exp(\eta'x)\Bigg(\sum_{i=2}^{k}\exp(\beta(t_{i} - t_{i-1})) - (k - 1)\Bigg),
\end{equation*}
\begin{equation*}
b = \frac{\lambda}{\beta}\exp(\eta'x)\Bigg(\sum_{i=2}^{k}\exp(\beta(t_{i} - t_{i-1})) + \exp(\beta t_{k+1} - \beta t_{k}) - k\Bigg),
\end{equation*}
and, $u \sim U(0, 1)$.

Now consider $ t_{k} + t_{s} \leq t < t_{k+1} $ in a multiple-dose setting, where all infusions up till the $k^{th}$ perfectly adhere to the 8-weekly schedule. The cumulative hazard function is equal to
\begin{eqnarray}
H(t, x, z(t)) &=& \int_{0}^{t}\exp(\beta z(u) + \eta'x) du \nonumber\\
&=& \lambda\exp( \eta'x)\int_{0}^{t}\exp(\beta z(u)) du\nonumber\\
&=& \lambda\exp( \eta'x)\Bigg[\int_{0}^{t_{k}}\exp(\beta z(u)) du + \int_{t_{k}}^{t_{k} + t_{s}}\exp(\beta(u - t_{k}) du 
+ \int_{t_{k} + t_{s}}^{t}\exp(\beta(t_{s})) du\Bigg]\nonumber\\
&=& \frac{\lambda}{\beta}\exp(\eta'x)\Bigg[\sum_{i=2}^{k} \exp(\beta(t_{i}-t_{i-1})) + \exp(\beta(t_{s})) + \beta(t - t_{s} - t_{k})\exp(\beta(t_{s})) - k\Bigg].\nonumber
\end{eqnarray}
And, the inverse cumulative function is
\begin{equation*}
H^{-1}(u) = \frac{1}{\beta\exp(\beta t_{s})}\Bigg(\frac{\beta u}{\lambda\exp(\eta'x)} - \sum_{i=2}^{k} \exp(\beta(t_{i} - t_{i-1})) - \exp(\beta t_{s})  + k \Bigg) + t_{s} + t_{k}.
\end{equation*}
Therefore, an event time can be generated as
\begin{multline}
T = \frac{1}{\beta\exp(\beta t_{s})}\Bigg(\frac{\beta(-\log(u))}{\lambda\exp(\eta'x)} - \sum_{i=2}^{k} \exp(\beta(t_{i} - t_{i-1})) - \exp(\beta t_{s})  + k \Bigg) + t_{s} + t_{k}\mbox{, if}\\
a \leq -\log(u) < b,
\end{multline}
where, 
\begin{eqnarray}\nonumber
a &=&  \frac{\lambda}{\beta}\exp(\eta'x)\Bigg[\sum_{i=2}^{k} \exp(\beta(t_{i}-t_{i-1})) + \exp(\beta t_{s}) - k\Bigg],\nonumber\\
b &=& \frac{\lambda}{\beta}\exp(\eta'x)\Bigg[\sum_{i=2}^{k}\exp(\beta(t_{i} - t_{i-1})) + \exp(\beta t_{k+1} - \beta t_{k}) - k\Bigg],\nonumber
\end{eqnarray}
and, $u \sim U(0, 1)$. \\

If more infusions continue to be given after a violation of the infusion schedule, and $ t_{k+n} \leq t < t_{k+n} + t_{s}$ and $t_{k} \leq t_{k} + t_{s} < t_{k+1}$, then the cumulative hazard function is equal to 
\begin{align}
H(t, x, z(t)) =& \int_{0}^{t}\exp(\beta z(u) + \eta'x) du\nonumber\\
=& \lambda\exp( \eta'x)\int_{0}^{t}\exp(\beta z(u)) du\nonumber\\
=&\lambda\exp(\eta'x)\Bigg(\int_{0}^{t_{k}}\exp(\beta z(u) du+\int_{t_{k}}^{t_{k}+ t_{s}}\exp(\beta(u - t_{k}) du \nonumber\\
&+ \int_{t_{k} + t_{s}}^{t_{k+1}}\exp(\beta(t_{s})) du + \int_{t_{k+1}}^{t}\exp(\beta z(u)) du\Bigg)\nonumber\\
=& \frac{\lambda}{\beta}\exp(\eta'x)\Bigg[\sum_{i=2}^{k} \exp(\beta(t_{i}-t_{i-1})) + \sum_{i=k+2}^{k+n} \exp(\beta(t_{i}-t_{i-1})) \nonumber\\
&+ \exp(\beta t - \beta t_{k+n}) + \beta(t_{k+1} - t_{s} - t_{k})\exp(\beta t_{s}) +  \exp(\beta t_{s})  - (k+n)\Bigg].\nonumber
\end{align}

And, the inverse cumulative function is
\begin{multline}
H^{-1}(u) = \frac{1}{\beta}\log\Bigg(\exp(\beta t_{k+n})\Bigg(\frac{\beta(u)}{\lambda\exp(\eta'x)} - \sum_{i=2}^{k} \exp(\beta(t_{i} - t_{i-1})) - \sum_{i=k+2}^{k+n} \exp(\beta(t_{i} - t_{i-1}))\\
 - \beta(t_{k+1} - t_{s} - t_{k})\exp(\beta t_{s}) -  \exp(\beta t_{s})  + (k+n) \Bigg)\Bigg).\nonumber
\end{multline}
Therefore, an event time can be generated as
\begin{multline}
T = \frac{1}{\beta}\log\Bigg(\exp(\beta t_{k+n})\Bigg(\frac{\beta(-\log(u))}{\lambda\exp(\eta'x)} - \sum_{i=2}^{k} \exp(\beta(t_{i} - t_{i-1})) - \sum_{i=k+2}^{k+n} \exp(\beta(t_{i} - t_{i-1}))\\
 - \beta(t_{k+1} - t_{s} - t_{k})\exp(\beta t_{s}) -  \exp(\beta t_{s})  + (k+n) \Bigg)\Bigg)\mbox{, if } a \leq -\log(u) < b  \mbox{, where}\\ 
 a = \frac{\lambda}{\beta}\exp(\eta'x)\Bigg[\sum_{i=2}^{k} \exp(\beta(t_{i}-t_{i-1})) + \sum_{i=k+2}^{k+n} \exp(\beta(t_{i}-t_{i-1})) \\
+ \beta(t_{k+1} - t_{s} - t_{k})\exp(\beta t_{s}) +  \exp(\beta t_{s})  - (k+n - 1)\Bigg],\\
b = \frac{\lambda}{\beta}\exp(\eta'x)\Bigg[\sum_{i=2}^{k} \exp(\beta(t_{i}-t_{i-1})) + \sum_{i=k+2}^{k+n} \exp(\beta(t_{i}-t_{i-1})) \\
+ \exp(\beta t_{k+n+1} - \beta t_{k+n}) + \beta(t_{k+1} - t_{s} - t_{k})\exp(\beta t_{s}) +  \exp(\beta t_{s})  - (k+n)\Bigg],\nonumber
\end{multline}
and $u \sim U(0, 1)$. 

The strategies described above can be extrapolated to settings where multiple violations to the 8-weekly infusion schedule occur.

\end{document}